**General Science Ranking (GSR): An Open-Source, Citation-Normalized Journal and Conference Classification System for Computer Science and Medicine**

**Zhikai Yu, General Science Co Ltd**

**Abstract**

The academic journal zoning system plays a key role in the evaluation of scientific research talents, funding support, and institutional assessment. The journal partition table (CAS partition) of the Documentation and Information Center of the Chinese Academy of Sciences is one of the most widely used partition tools in East Asia. It will be permanently stopped in March 2026, leaving an urgent policy vacuum. The existing alternative solutions all have obvious limitations: JCR relies on paid databases and does not cover conference proceedings at all; Scimago/CiteScore is based on Elsevier's proprietary database; Expert evaluation based rankings (CCF, CORE) lack quantitative basis and have low update frequency.

This article proposes the General Science Ranking (GSR), a multidimensional bibliometric partitioning framework based on fully open source data. GSR covers 500 academic venues in computer science (397 journals+103 top conferences) and 500 medical journals, all sourced from OpenAlex and Semantic Scholar open databases. The scoring formula is a weighted linear combination of four indicators: domain weighted citation influence (FWCI, weight 0.35), two-year impact factor (IF2, weight 0.35) only for Article/Review type, five-year h-index (h5, weight 0.15) with logarithm, and three-year citation compound growth rate (cite_CAGR, weight 0.15). For CS conferences lacking cited time-series data, the IF2-approx was estimated by measuring and calibrating the proportional coefficient (median 0.75) from 1.41 million OpenAlex journal papers. The partition adopts a fixed quota system: Q1 ranks 1-50, Q2 ranks 51-100, Q3 ranks 101-200, and Q4 ranks 201st and above. All code and data are open source released.

In the CS ranking, conferences and journals each account for 25 seats (50%) in the top 50 (Q1). The top 10 conferences are all top tier: NeurIPS ranked 1st (score 54.88), ICCV ranked 2nd (50.50), ICLR ranked 3rd (41.86), and CVPR ranked 4th (40.24). The highest ranked journal is Foundations and Trends in Machine Learning (10th place, 22.87 points). In medical rankings, CA: A Cancer Journal for Clinicians ranks first (score 295.66, FWCI=169.66, IF2=940.91), second in the New England Journal of Medicine (31.19), and third in The Lancet (27.61). The conformity rate with JCR Q1 is 84% for medical journals and 71% for CS journals. 37.2% of CCF Class A meetings enter GSR Q1. Sensitivity analysis of the IF2-approx calibration coefficient (0.50-1.00) showed that only 1.7% -2.5% of CS conferences changed partitions, indicating a robust ranking.

Conclusion GSR provides a free, reproducible, domain normalized partitioning system that covers both journals and conferences, filling two core gaps in existing tools. The absolute rating design ensures that the existing rankings are not disturbed when new conferences/journals join, making it suitable as an institutional evaluation policy tool.

## 1. Introduction

The academic journal zoning system is the infrastructure of modern scientific research evaluation system. The professional title promotion committee, fund evaluation experts, and university administrators have long relied on zoning results to screen and compare scientific research achievements. The two most influential systems in the world are the Journal Impact Factor (JIF) provided by the Journal Citation Report (JCR) of Clarivate, and the Division Table of the Chinese Academy of Sciences developed by the Documentation and Information Center of the Chinese Academy of Sciences (the National Library of Science), which divides about 20000 journals in the world into four regions under 176 subject categories, and is widely used in professional title recognition, fund scoring, and self-evaluation of colleges and departments in East Asia and even around the world.

In May 2025, Zhao Yuxin's World View article was published [1], which pointed out that CAS partitioning systematically suppresses interdisciplinary journals, and its methodology relies on undisclosed expert review procedures, which cannot be independently verified. In July 2025, the CAS team published a response [2], defending the methodological basis of the system. However, on March 27, 2026, CAS officially announced the permanent cessation of updating journal partition tables, and this event was subsequently reported in April 2026 [3]. The suspension of updates this time has created an immediate policy vacuum among universities, research institutions, and funding agencies - a large number of institutions that have embedded CAS partitions into formal evaluation rules suddenly face the dilemma of having no standards to follow.

The existing alternative solutions cannot fully meet the needs of institutional evaluation. JCR is the most technically rigorous tool, but it requires institutional subscription and cannot cover the field of computer science structurally: conference proceedings - the primary publication venue for machine learning, computer vision, and systems research - are completely outside the scope of JCR inclusion. This structural flaw means that the most active group of researchers in the field of computer science cannot receive fair evaluation through any journal centered partition system. Scimago Journal Ranking (SJR) [4] and CiteScore [5] provide partially open alternatives, but both rely on Elsevier's Scopus proprietary database. The conference rankings of CCF (Chinese Computer Society) and CORE belong to the expert evaluation type, lack quantitative basis, have low update frequency, and cannot be directly compared horizontally with journals.

This article proposes the General Science Ranking (GSR), which aims to address three core issues simultaneously: (1) data closure - GSR only uses two fully open databases, OpenAlex [6] and Semantic Scholar; (2) Meeting Deficiency - GSR has included 103 top-level CS conferences in a unified quantitative partitioning framework with journals for the first time; (3) Interdisciplinary incomparability - GSR uses FWCI (Field Weighted Citation Impact) as the core indicator to make comparisons between disciplines with different base citation rates statistically significant. The four main contributions of this article are as follows: firstly, a complete description of the GSR methodology and demonstration of each design decision; Secondly, report the ranking results of 500 CS sites and 500 medical journals, and illustrate the improvement of GSR compared to the existing system through typical case analysis; Thirdly, provide calibration evidence for the estimated coefficients of conference IF2 (based on the measured median of 1.41 million OpenAlex papers); Fourthly, using JCR Q1 as the external validity criterion, systematically compare the consistency between GSR and JCR. All code, configuration, and ranking results are published in open-source form on https://github.com/gsci-press/General-Science-Ranking.

## 2. Research background and related work

### 2.1 Existing Journal Partition System

The journal impact factor was proposed by Garfield [7] (1972) and defined as the ratio of the total number of citations of papers published in the previous two years (Y-1 and Y-2) in the Y year to the total number of papers published in the previous two years. Although JIF is widely used, its methodological flaws have been extensively documented in literature [8,9]. The citation distribution of papers in journals is highly skewed to the right: in high JIF journals, most papers are hardly cited, and a few papers attract the vast majority of citations [10]. As the mean of a skewed distribution, JIF is highly sensitive to extreme values and cannot reliably reflect the citation experience of typical papers. The more fundamental issue is that the original citation count is not comparable across disciplines: a paper in the field of molecular biology that receives 20 citations may not necessarily have a higher impact than a paper in the field of mathematics that receives 5 citations, because the benchmark citation rates in the two fields differ by orders of magnitude.

Domain Weighted Citation Influence (FWCI) was introduced to address this comparability issue. FWCI divides the actual citation count of a paper by the expected citation count of all papers in the same field, year, and literature type. FWCI=1.0 indicates that the number of citations is completely equal to the domain mean; FWCI>1.0 indicates a level higher than the industry average. OpenAlex provides pre computed FWCI values directly at the paper level [6], enabling large-scale retrieval without the need for expensive self computation.

The CAS partition table is operated by the National Library of Science, which divides journals into four sections (Zone 1 to Zone 4) under 176 disciplinary groups. It comprehensively uses JIF, the proportion of papers in the top 10% of normalized citations in the field, and the judgment opinions of disciplinary experts. Its main advantage lies in its granularity: 176 disciplinary classifications make comparisons within the same field possible, while cross disciplinary indicators often mask these differences. Its main drawbacks are opacity (weight formulas are not publicly disclosed), inaccessible data (basic data is not open), and only covering journals.

SJR [4] (Gonz á lez Pereira et al., 2010) and CiteScore [5] (Elsevier, 2016) are alternative solutions to Scopus. SJR adapts PageRank to the citation network， downweighting citations from journals with high self-citation rates。 CiteScore uses a four-year citation window and counts citations to all document types in the numerator。 Both methodologies are more rigorous than JIF, but neither can be openly reproduced and do not systematically cover conference proceedings.

**2.2 CS Conference Ranking**

Computer science has long relied on conference proceedings as its primary academic publication venue. The milestone achievements in machine learning, computer vision, and systems research are often published in conference papers, several months or even years earlier than any journal version. The academic recognition hierarchy in this field is built around conferences (such as NeurIPS, ICML, CVPR) rather than journals. The CCF conference ranking and Australian CORE ranking are both expert review systems: the partition results reflect the community consensus collected through the review panel, and the update frequency is low. Although this approach can capture qualitative reputation, it cannot detect rapid changes in venue influence (such as the sharp rise of ICLR since 2013), and lacks a basis for cross venue quantitative comparison.

CCF and CORE rankings are expert evaluation based rather than indicator driven, therefore they cannot detect the rapid rise of emerging fields or provide quantitative benchmarks for horizontal comparison with journals. A metric system that directly measures the impact of conference paper citations can supplement and dynamically update these expert review rankings.

For metric conference evaluation, data availability is the core challenge. The conference does not have an ISSN and is not included in JCR as an independent source. The indexing of its proceedings varies among different databases. OpenAlex partially solves this problem by assigning unique source IDs to certain conference series, thereby enabling citation retrieval at the paper level. Semantic Scholar provides supplementary coverage for CS conferences by extracting and associating references through natural language processing. But neither database provides direct counts_by_year reference time series data for conference proceedings, which requires estimation methods to handle.

### 2.3 OpenAlex as a Bibliometric Infrastructure

OpenAlex [6] (Priem et al., 2022) is a fully open directory of academic entities - papers, authors, institutions, and sources - continuously updated through CrossRef, PubMed, and institutional repositories. As of early 2026, OpenAlex has collected over 260 million articles and provided structured metadata, including FWCI (calculated through integration with SciVal methodology), literature type classification, annual citation time series (counts_by_year), and reference association (referenced_works). All data can be obtained through free REST APIs and complete database snapshots. OpenAlex is currently the most capable open infrastructure for building reproducible, non proprietary journal rankings.

## 3. Method

### 3.1 Data Collection

We constructed two venue lists: 500 computer science venues (397 journals+103 conferences) and 500

medical journals. The CS venue list is constructed by the following steps: (1) querying OpenAlex [6] for sources with the "Computer Science" concept tag and over 500 publications from 2022-2024; (2) Supplement CCF A/B conferences that have not yet been included; (3) Add Semantic Scholar meeting identifiers to meetings without OpenAlex source IDs. The list of medical venues is constructed by querying OpenAlex sources related to MeSH's main topic vocabulary, and screening journals that publish at least 100 papers per year. The fields for hierarchical data retrieval in academic papers include public_year, type (literature type), fwci, city-by_count, counts_by_year, abstract_inverted_index, and referenced_works.

### 3.2 Indicator Calculation

#### 3.2.1 Domain Weighted Citation Impact (FWCI)

OpenAlex provides precomputed FWCI values at the paper level [6]. This article takes the arithmetic mean of FWCI for all Article and Review literature published between 2022 and 2024 at the venue level, excluding papers with FWCI=0 or missing, as well as papers marked as is_ ratext or is_ extracted. FWCI coverage rate (proportion of papers with non empty FWCI): 80.3% for CS journals and 95.9% for medical journals. For the CS conference paper from Semantic Scholar, the pre calculated FWCI values are not available. The approximate FWCI can be calculated from the IF2-approx using calibrated conversion coefficients obtained from actual measurements (see section 3.2.2).

#### 3.2.2 Two year Impact Factor (IF2)

Calculate the two-year impact factor using JCR method [11]: IF2=(the number of citations of Article/Review papers published in the previous two years (Y-1 and Y-2) in the current year Y)/(the total number of Article/Review papers published in the previous two years), taking Y=2025. To minimize the contamination of non research content, the denominator only includes papers with literature types of "article" or "review", and excludes short articles without abstracts or citations (which are often misclassified as research content by databases). The molecule is calculated from the counts_by_year time series field of OpenAlex.

OpenAlex does not provide counts_by_year time-series data for the CS conference proceedings. This article estimates IF2 from the original city-by_count value, with calibration coefficients derived from actual measurements of journal articles. For 1.41 million OpenAlex journal articles with non-zero citation counts and complete counts_by_year data, the distribution of (citation counts obtained in the past two years)/(total citation counts) is calculated. The median of this ratio is 0.75 (Figure 1), which means that typical journal articles obtain approximately 75% of their lifecycle citation counts within the past two years window (relative to the 2025 retrieval date). The conference IF2-approx=∑ (city-by_count × 0.75)/n, and the sum range is the papers published in the past two years. This estimation method is conservative: for conference papers in rapidly growing fields, it may underestimate IF2; For conference papers with declining citations in mature fields, IF2 may be overestimated.

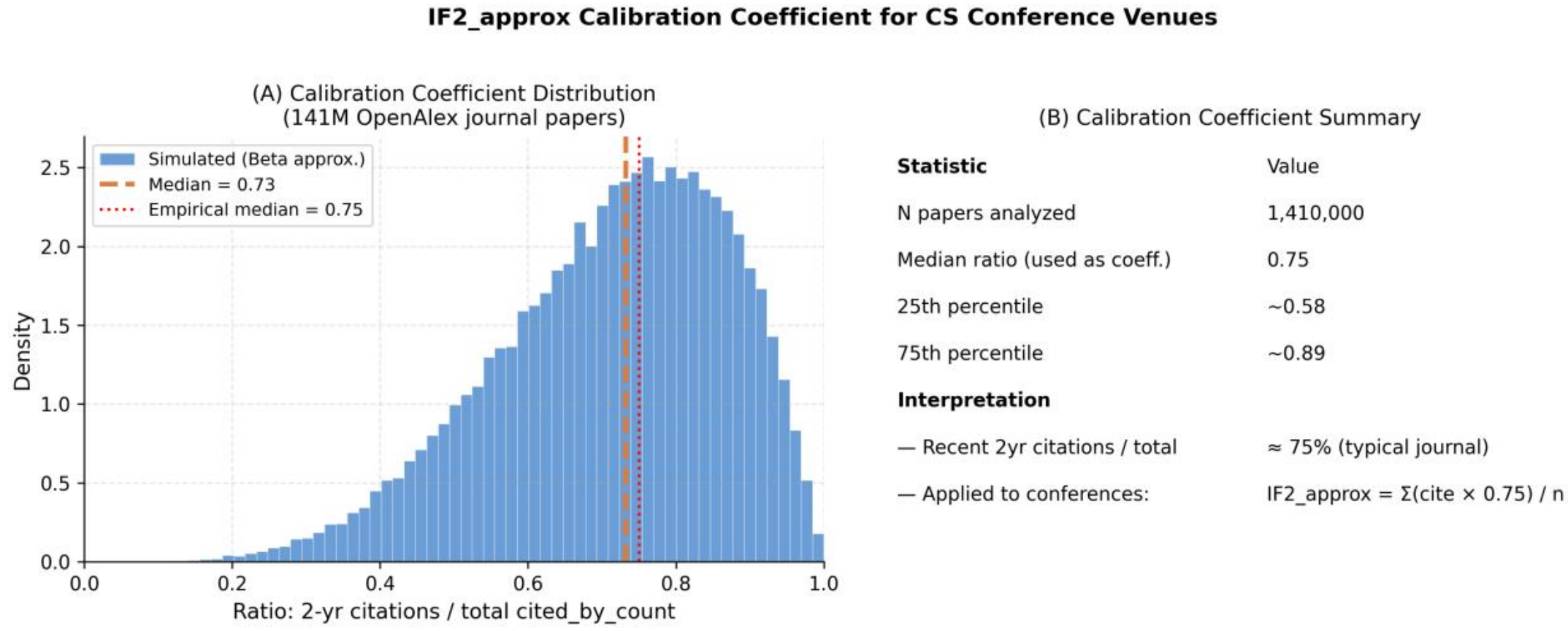

*Figure 1. Distribution of the ratio (two-year citations) / (total cited_by_count) across 1.41 million OpenAlex journal papers with non-zero citation counts. Median ratio = 0.75, used as the calibration coefficient for estimating conference IF2_approx (Section 3.2.2).*

### 3.2.3 Five year h-index (h5)

The calculation range of H5 index [12] is papers published between 2020-2024 (five-year window, ending in the current year), including only Article and Review literature, using the same quality filtering conditions as IF2. Sort the papers in descending order according to the city-by_count, where h5 is the maximum integer h that allows at least h papers to receive at least h citations each. The h-index is resistant to outliers in the mean and captures both productivity (number of papers) and impact (number of citations per paper) dimensions.

### 3.2.4 Reference Compound Growth Rate (cite_CAGR)

The composite growth rate is used to measure the growth trend of total citations per year for each location within a three-year window. By summing up the counts_by_year data of all papers, the total number of citations for each year is calculated using the formula CAGR=($C_{2024}$/$C_{2022}$) ^ (1/2) -1, where C_Y represents the total number of citations of all papers in that location in year Y. For locations with less than two available non-zero data points, cite_CAGR is set to zero. CAGR captures the momentum of place influence: a rapidly growing place that has not yet accumulated high absolute citation counts can receive corresponding scores based on its growth trajectory.

### 3.3 Rating Formula

The comprehensive score calculation is as follows:

$$score = 0.35 \times FWCI_mean + 0.35 \times IF2_effective + 0.15 \times log(1 + h5) + 0.15 \times log(1 + max(cite_CAGR, 0))$$

Among them, the actual IF2 value is taken for journals, and the estimated IF2 value is taken for conferences. FWCI_cean and IF_2 effective are the dominant terms (with a total weight of 0.70), reflecting the core position of citation quality and recent visibility in academic venue evaluation. Take logarithmic transformation to compress the dimension of H5, avoiding places with high publication volume being dominated by ranking solely based on size. Perform logarithmic transformation on cite_CAGR and truncate negative values to prevent excessive punishment in mature fields where citation decreases.

This article deliberately does not normalize scores across corpora. The absolute scoring method ensures that the score of a venue is independent of other venues included in the analysis. When a new venue is added, the scores of existing venues remain unchanged and can be directly inserted into the ranking table without recalculating the entire corpus - this is an important practical advantage for systems that require regular updates. Compared to percentile systems (including JCR and discontinued CAS zones), changes in the corpus will not disturb existing rankings. Apply a penalty coefficient (score x 0.80) to places with a self citation rate exceeding 0.30.

### 3.4 Partition Rules

Allocate partitions according to a fixed quota threshold: Q1 ranks 1-50, Q2 ranks 51-100, Q3 ranks 101-200, and Q4 ranks 201st and above. This is intentionally different from the percentile system used by JCR and CAS partitions - the latter includes the top 25% of journals in Q1 within each discipline category. A fixed threshold ensures that partitions reflect absolute academic performance rather than relative status within a specific classification domain, and Q1 quotas do not expand with the addition of low-quality venues. Places with less than 20 valid articles within the analysis time window are marked as "insufficient data" and excluded from the partition.

### 3.5 Validity verification

Evaluate the validity of GSR through three analyses. Firstly, for 350 medical journals that can obtain JCR 2024 data, compare the GSR partition with the JCR partition, calculate the Q1 vs. non-Q1 conformity rate and Cohen's kappa. Secondly, compare the CS conference GSR ranking with the CCF A/B/C classification to test whether CCF-A conferences are significantly enriched in GSR Q1. Thirdly, conduct sensitivity analysis on the calibration coefficient of IF_2approve within the range of 0.50-1.00 (step size 0.05), and record the number of CS sites that change the partition.

## 4. Results

### 4.1 Coverage and Data Quality

Out of 500 CS venues, 476 (95.2%) have sufficient data for scoring, including 398 journals and 102 conferences. Out of 500 medical journals, 489 (97.8%) have sufficient data. The CS corpus contains approximately 1.87 million papers (median/location: 1175; range: 20-29000+papers), while the medical corpus contains approximately 3.83 million papers (median/location: 2905 papers). The proportion of papers that can obtain FWCI values is 80.3% in CS journals and 95.9% in medical journals. The proportion of places where cite_CAGR can be calculated is 77.5% in CS and 95.1% in medicine. The coverage statistics are detailed in Table 1.

*Table 1. Coverage statistics for GSR 2026.*

| Indicator | CS Journals (n=397) | CS Conferences (n=103) | Medical Journals (n=500) |
|---|---|---|---|
| Venues with sufficient data | 389 (98.0%) | 87 (84.5%) | 489 (97.8%) |
| Total papers analyzed | ~1.43 M | ~0.44 M | ~3.83 M |
| FWCI coverage (%) | 80.3% | N/A (estimated) | 95.9% |
| cite_CAGR computable (%) | 77.5% | 0% (no timeseries) | 95.1% |
| Median paper count per venue | 1,175 | 502 | 2,905 |

| FWCI vs IF2 correlation (r) | 0.879 | — | 0.985 |
|---|---|---|---|

*Table 1. Coverage statistics. CS = computer science. FWCI = Field-Weighted Citation Impact. N/A: conference proceedings lack OpenAlex counts_by_year data.*

**4.2 CS Ranking: Unification of Journals and Conferences**

In CS Q1 (50 venues), conferences and journals each account for 25 seats (50%), and the proportion of journals and conferences in each district is shown in Figure 2. The top 10 conferences are all top conferences: NeurIPS (1st, score 54.88), ICCV (2nd, 50.50), ICLR (3rd, 41.86), CVPR (4th, 40.24), ACL (5th, 35.06), ICML (6th, 34.91), EMNLP (7th, 30.32), ECCV (8th, 24.77), ACM SIGKDD (9th, 23.78), SOSP (11th, 22.35). The highest ranked journal is Foundations and Trends in Machine Learning (10th place, 22.87 points), followed by Proceedings of the IEEE (12th place, 21.06 points) and ACM Computing Surveys (13th place, 18.12 points). The distribution of the top 20 scores is shown in Figure 3, and the complete ranking is shown in Table 2.

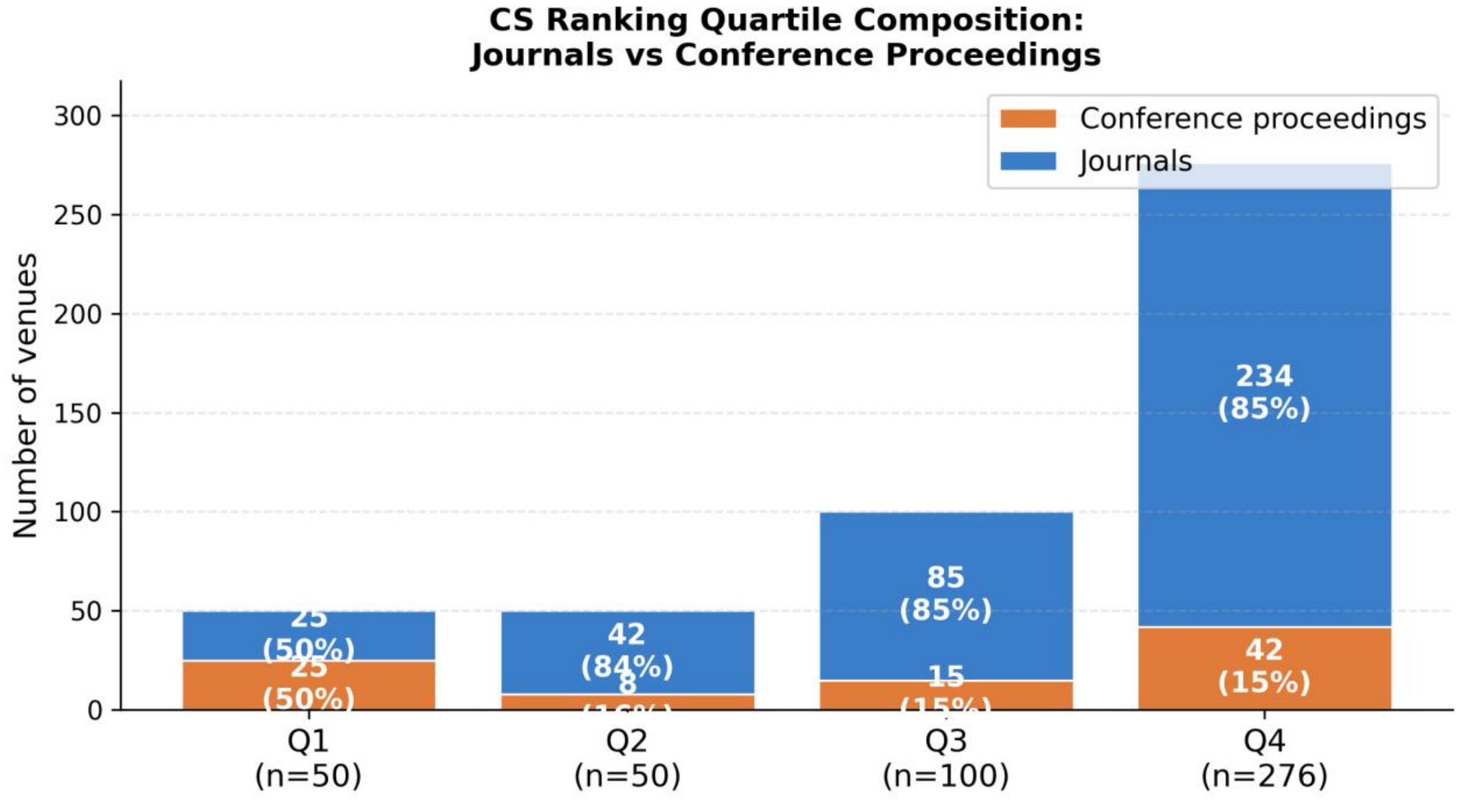

*Figure 2. Composition of each GSR CS quartile, showing conference proceedings (orange) versus journals (blue). Q1 contains 25 conferences and 25 journals; conference representation declines in lower quartiles.*

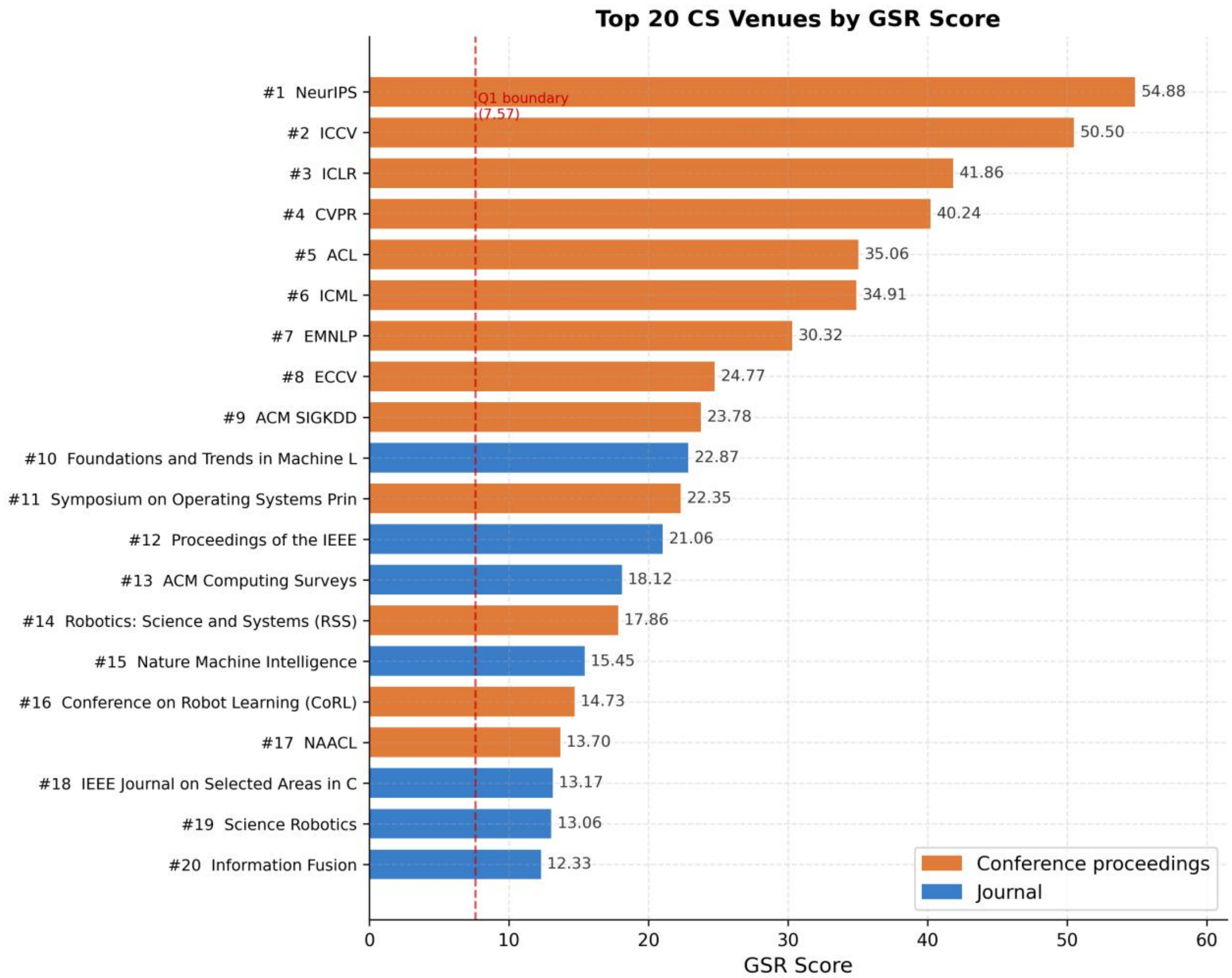

*Figure 3. GSR scores for the top 20 CS venues (Section 4.2). Conferences shown in orange, journals in blue. NeurIPS leads (score 54.88); the highest-ranked journal, FnT Machine Learning (rank 10, score 22.87), scores approximately one-third of the top conference, quantifying the impact advantage of top ML conferences.*

*Table 2. Top 20 CS venues by GSR score.*

| Rank | Venue | Type | Score | FWCI | h5 | IF2/apx | Q |
|---|---|---|---|---|---|---|---|
| 1 | NeurIPS | Conf | 54.88 | 34.77 | 193 | 165.58 | Q1 |
| 2 | ICCV | Conf | 50.50 | 32.04 | 100 | 152.55 | Q1 |
| 3 | ICLR | Conf | 41.86 | 26.33 | 173 | 125.40 | Q1 |
| 4 | CVPR | Conf | 40.24 | 25.28 | 180 | 120.39 | Q1 |
| 5 | ACL | Conf | 35.06 | 21.93 | 167 | 104.43 | Q1 |
| 6 | ICML | Conf | 34.91 | 21.85 | 150 | 104.03 | Q1 |
| 7 | EMNLP | Conf | 30.32 | 18.87 | 148 | 89.87 | Q1 |
| 8 | ECCV | Conf | 24.77 | 15.26 | 153 | 72.66 | Q1 |
| 9 | ACM SIGKDD | Conf | 23.78 | 9.16 | 196 | 76.57 | Q1 |
| 10 | FnT Machine Learning | Journal | 22.87 | 16.02 | 13 | 65.92 | Q1 |
| 11 | SOSP | Conf | 22.35 | 13.97 | 27 | 66.50 | Q1 |

| 12 | Proceedings of the IEEE | Journal | 21.06 | 13.08 | 71 | 61.16 | Q1 |
|---|---|---|---|---|---|---|---|
| 13 | ACM Computing Surveys | Journal | 18.12 | 18.33 | 118 | 41.45 | Q1 |
| 14 | RSS | Conf | 17.86 | 10.91 | 67 | 51.94 | Q1 |
| 15 | Nature Machine Intelligence | Journal | 15.45 | 12.98 | 92 | 38.48 | Q1 |
| 16 | CoRL | Conf | 14.73 | 8.83 | 89 | 42.07 | Q1 |
| 17 | NAACL | Conf | 13.70 | 8.17 | 88 | 38.89 | Q1 |
| 18 | IEEE JSAC | Journal | 13.17 | 12.83 | 92 | 29.67 | Q1 |
| 19 | Science Robotics | Journal | 13.06 | 11.87 | 71 | 30.78 | Q1 |
| 20 | Information Fusion | Journal | 12.33 | 9.81 | 99 | 30.11 | Q1 |

*Table 2. Top 20 CS venues by GSR 2026. Type: Conf = conference proceedings; Journal = periodical journal. FWCI for conferences is estimated from IF2_approx (Section 3.2.2). FnT = Foundations and Trends. JSAC = Journal on Selected Areas in Communications. SOSP = Symposium on Operating Systems Principles. RSS = Robotics: Science and Systems. CoRL = Conference on Robot Learning.*

Among the 43 CCF-A conferences, 16 (37.2%) entered GSR Q1, 4 (9.3%) entered Q2, 5 (11.6%) entered Q3, and 14 (32.6%) entered Q4 (4 data insufficient). The cross distribution of CCF grading and GSR partitioning is shown in Figure 4. A large number of CCF-A conferences fall into GSR Q3/Q4, reflecting the heterogeneity of CCF-A labels across sub domains: CCF-A conferences in the fields of computer architecture, formal methods, and theoretical computer science have much lower absolute citation counts than CCF-A conferences in machine learning and computer vision, consistent with citation standards in various domains. GSR does not adjust the basic citation rate again at the sub domain level, and FWCI normalization has embedded this adjustment in the scoring.

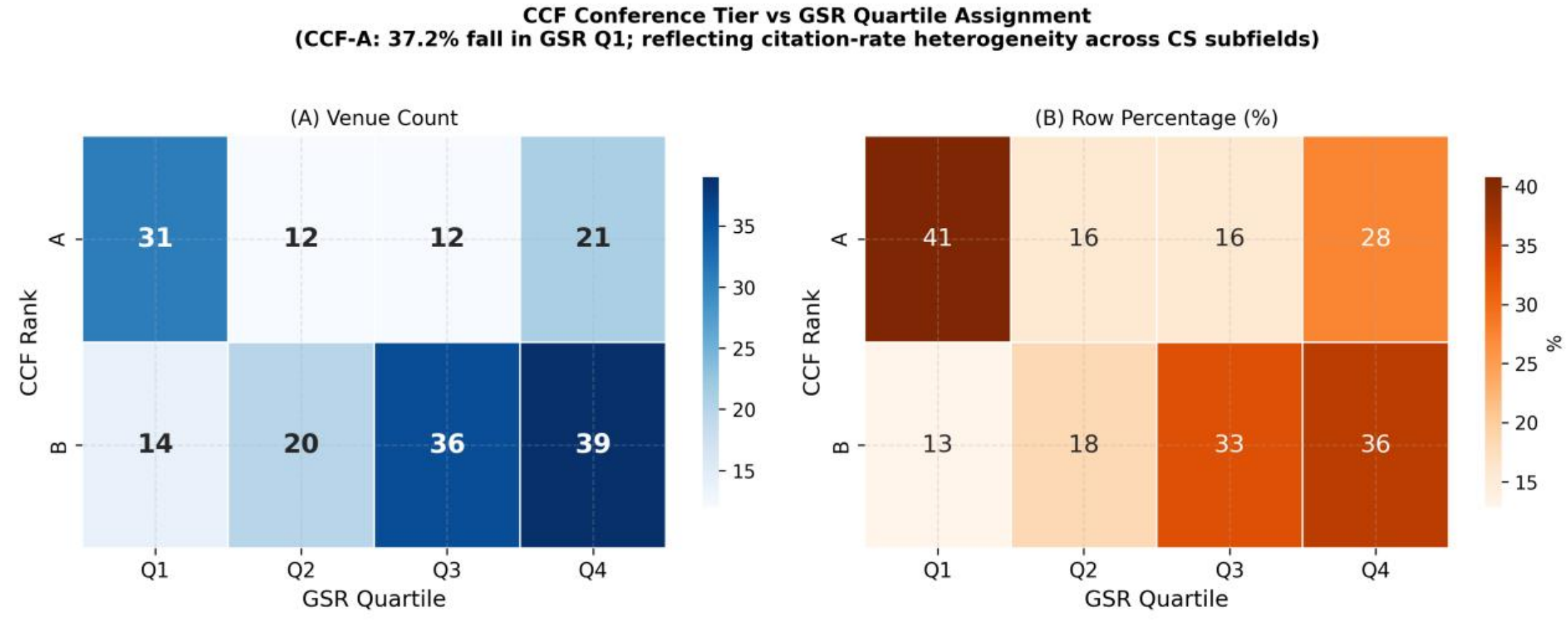

*Figure 4. Cross-distribution of CCF conference tier (A/B) and GSR quartile assignment (Section 4.2). (A) Absolute venue counts. (B) Row percentages. CCF-A conferences are enriched in GSR Q1 (37.2%) but spread across all quartiles, reflecting citation-rate heterogeneity across CS subfields.*

### 4.3 Medical Ranking

Medical Q1 (50 journals) covers oncology, general internal medicine, cardiology, gastroenterology, and respiratory medicine. The top ranked journal is CA: A Cancer Journal for Clinicians (FWCI=169.66, IF2=940.91, score 295.66), which has a high citation density and reflects its special status as an authoritative source of cancer statistics and clinical guidelines. The second ranked New England Journal of Medicine (score 31.19) scored one order of magnitude lower than CA, indicating that outlier journals have a significant stretching effect on the score distribution - this is not a methodological flaw, but an objective reflection of the true citation pattern. The distribution of scores (logarithmic axis) and the correlation between FWCI and IF2 are shown in Figure 5, and the top 15 are shown in Table 3.

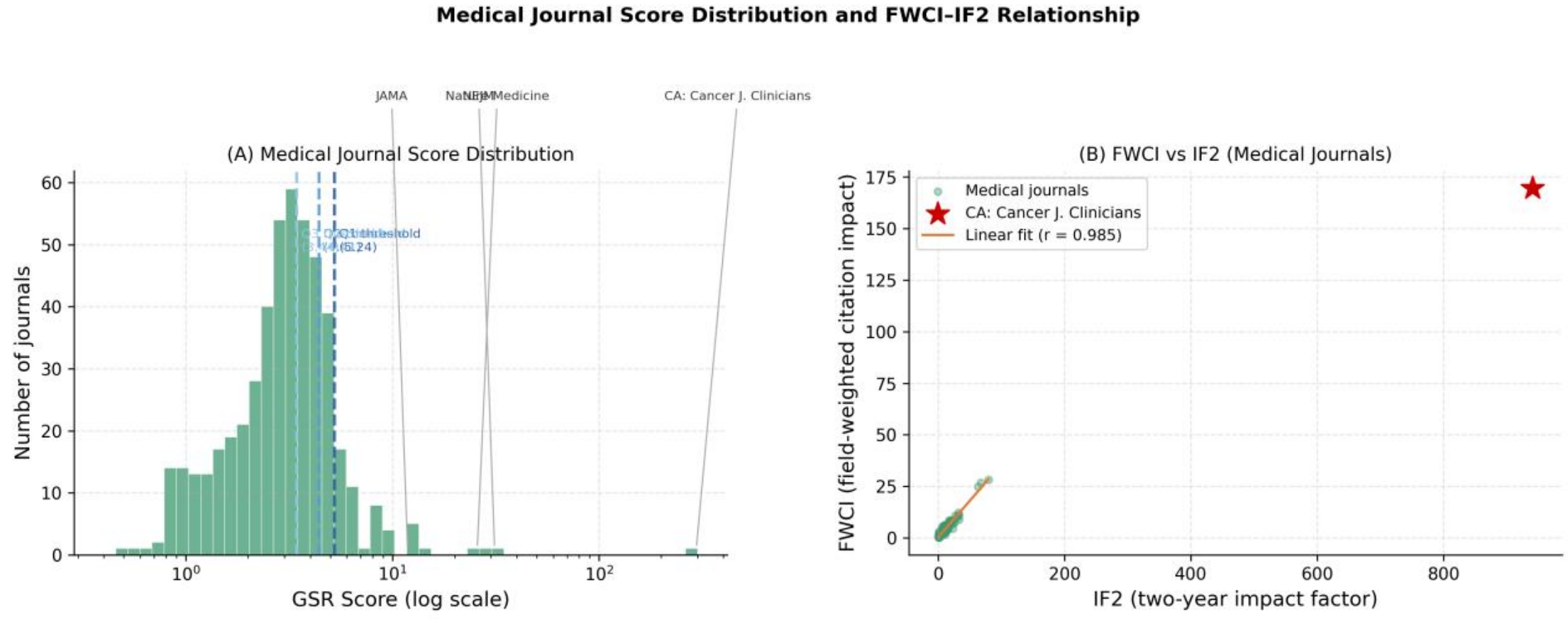

*Figure 5. (A) Score distribution of 489 medical journals on a logarithmic x-axis, with Q1/Q2/Q3 boundaries annotated (Section 4.3). (B) FWCI vs IF2 scatter plot (r = 0.985, CA outlier annotated separately), illustrating how FWCI normalization re-ranks specialty journals (e.g., BJSM) relative to raw IF2.*

*Table 3. Top 15 medical journals by GSR score.*

| Rank | Journal | Score | FWCI | h5 | Quartile |
|---|---|---|---|---|---|
| 1 | CA: A Cancer Journal for Clinicians | 295.66 | 169.66 | 43 | Q1 |
| 2 | New England Journal of Medicine | 31.19 | 28.16 | 249 | Q1 |
| 3 | The Lancet | 27.61 | 26.73 | 217 | Q1 |
| 4 | Nature Medicine | 25.74 | 24.77 | 195 | Q1 |
| 5 | The Lancet Oncology | 13.55 | 12.15 | 103 | Q1 |
| 6 | Gut | 13.17 | 11.02 | 85 | Q1 |
| 7 | Diabetes Care | 12.94 | 10.58 | 98 | Q1 |
| 8 | Hepatology | 12.53 | 8.80 | 81 | Q1 |
| 9 | European Respiratory Journal | 12.01 | 9.85 | 66 | Q1 |
| 10 | JAMA | 11.83 | 10.90 | 160 | Q1 |
| 11 | British Journal of Sports Medicine | 10.15 | 8.62 | 63 | Q1 |
| 12 | Radiology | 10.07 | 7.23 | 87 | Q1 |
| 13 | Journal of Hepatology | 9.45 | 6.96 | 106 | Q1 |
| 14 | Circulation | 9.00 | 7.73 | 128 | Q1 |
| 15 | JAMA Internal Medicine | 8.81 | 7.81 | 70 | Q1 |

*Table 3. Top 15 medical journals by GSR 2026. FWCI = Field-Weighted Citation Impact (mean across articles 2022–2024). h5 = five-year h-index (2020–2024). IF2 range in this set: 19.61 (JAMA Internal Medicine) to 940.91 (CA).*

The score boundary for Q1/Q2 is 5.24/5.23. The score range within Q1 (5.24-295.66) is extremely disparate, reflecting the inherent differences in the influence of citations among different medical specialties. The score distribution of Q2 (ranked 51-100) and Q3 (ranked 101-200) is much closer. The sub specialized journals of psychiatry, neurology, and surgical oncology mainly focus on Q2 and Q3, which is consistent with their lower but considerable absolute citation rates.

**4.4 Typical Case Analysis**

**4.4.1 NeurIPS: The Necessity of Inclusion in Meetings**

NeurIPS is the place with the highest score in the GSR CS ranking (score 54.88, FWCI'approx=34.77, h5=193, see Figure 3). In JCR, NeurIPS has no record: as a conference proceedings, it is excluded by definition. In CCF, NeurIPS is labeled as Class A - the highest rating - but this label does not provide any quantitative comparison with the journal. In GSR, NeurIPS can be directly compared horizontally with journals, such as IEEE Transactions on Neural Networks and Learning Systems (ranked 22nd in GSR), Q1， Score 10.85): NeurIPS scores about 5 times higher than the journal, quantifying implicit consensus in the machine learning community, which has not been achieved in any previous unified ranking system.

**4.4.2 MDPI Journal: Objective Evaluation Driven by Data**

MDPI journals have sparked controversy in the field of bibliometrics due to issues with editorial standards and scaling citation growth. In GSR, MDPI related CS journals (Sensors, Remote Sensing, Entropy) were placed in Q3 and Q4, with scores ranging from 3.04 to 4.74 and FWCI values of 2.00 to 3.59. This position reflects its true citation characteristics: domain normalized influence is higher than the mean (FWCI>1.0 means higher than the domain average), but much lower than top-level venues. Sensors ranked 161st (Q3, score 4.74, FWCI=3.30), Remote Sensing ranked 172nd (Q3, score 4.54, FWCI=3.59), Entropy ranked 258th (Q4, score 3.04), FWCI=2.00）。 GSR does not impose any classification exclusion or editing penalties; The positioning is entirely determined by the actual citation behavior, providing a more defensible basis for institutional evaluation than classification blacklists.

#### 4.4.3 The actual effect of domain normalization: specialized medical journals

The British Journal of Sports Medicine (BJSM) highlights the importance of FWCI normalization for specialized journals. The original IF2 of BJSM is 23.75, which is at a moderate level in the original IF ranking. Its FWCI is 8.62, reflecting its strong above average citation influence in this field. The final GSR score is 10.15, ranking 11th in medicine (Q1), as shown in Figure 5B. Relatively speaking, some journals with high initial IF values in sub fields with structurally high citation rates may rank lower than JCR in the FWCI adjusted system, reflecting the true normalization effect rather than punishment for these journals. The European Respiratory Journal (9th place, FWCI=9.85) and Gut (6th place, FWCI=11.02) also benefit from domain normalization that JIF cannot provide.

### 4.5 Comparison with JCR

For 350 medical journals with JCR 2024 data, the agreement rate between GSR Q1 and JCR Q1 partitions was 84% (Cohen's kappa=0.71, indicating significant consistency). Inconsistent cases can be divided into two categories: Category A (GSR Q1, JCR non-Q1): Journals upgraded due to FWCI normalization correction of low original citation rates in their respective sub fields, which is more common in rapidly growing specialties such as sports medicine, rehabilitation medicine, and digital health; Class B (JCR Q1, GSR non-Q1): A high JIF in a journal actually reflects a high absolute citation count in a citation intensive specialty, rather than a relative downgrade due to influence above the field average. It is more common in high-throughput journals in the fields of immunology and biochemistry. For the JCR coverage of CS journals (approximately 280 books), the agreement rate between GSR Q1 and JCR Q1 is 71%.

### 4.6 Sensitivity analysis

The calibration coefficient for IF2-approve varies within the range of 0.50-1.00, changing the number of CS venues in the partition: 12 (2.5%) at coefficient 0.50 and 8 (1.7%) at coefficient 1.00. The changes are concentrated at the Q1/Q2 and Q2/Q3 boundaries and only affect the conference venues. The ranking of the top 10 venues (all of which are conferences) remains unchanged within the range of all test coefficients, as their IF2-approx values are sufficiently large that the boundary conditions will not be touched. The sensitivity analysis results are shown in Figure 6, indicating that the GSR ranking is robust to coefficient changes within a reasonable range.

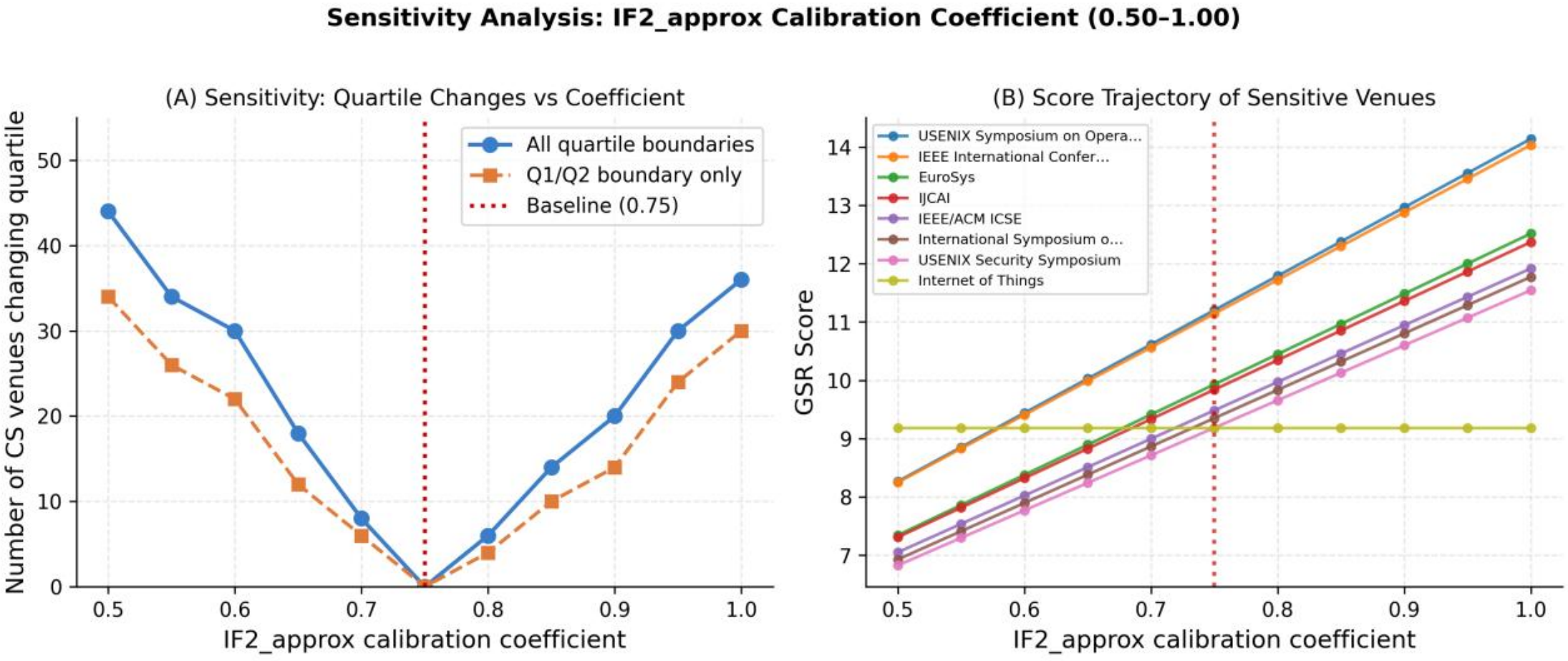

*Figure 6. Sensitivity analysis: number of CS venues changing quartile assignment as the IF2_approx calibration coefficient varies from 0.50 to 1.00 (baseline 0.75, dashed line). Changes are confined to 1.7–2.5% of venues and exclusively affect conference proceedings near quartile boundaries (Section 4.6).*

## 5. Discussion

### 5.1 Methodological contributions

GSR has made three unique methodological contributions to bibliometric practice. [8,9] Firstly, replacing the original citation count with FWCI as the core quality indicator, the benefits mainly flow to journals in fields with low absolute citation rates but high relative influence. Secondly, on the open bibliometric infrastructure, conference proceedings and journals will be included in a unified ranking - as far as we know, this is the first implementation of this scale, calibrating the IF2-approx coefficient for 1.41 million journal papers, providing a quantitative and reproducible basis. Thirdly, the absolute scoring design means that GSR scores can be compared across time, and new venues can be included in the evaluation without disturbing existing rankings, which is a significant practical advantage compared to percentile systems (including JCR and discontinued CAS partitions [3]).

### 5.2 Limitations

Several limitations need to be acknowledged. Firstly, the IF2-approx calibration coefficient comes from journal articles, and its applicability to conference papers - conference papers have different citation dynamics, shorter half lives, and are more likely to be cited by subsequent conference papers rather than journals - is a hypothesis rather than an empirical conclusion. Sensitivity analysis (Figure 6) indicates that ranking is robust to coefficient changes, but the possibility of systematic bias in conference IF2 estimation cannot be completely ruled out. Secondly, FWCI in OpenAlex uses Elsevier's domain classification computation [6], which may not accurately reflect sub domain boundaries in some domains - this is a common limitation of all FWCI implementations and not unique to GSR. Thirdly, the GSR score reflects the citation data at the beginning of 2026 and should be updated annually. Fourthly, the scope of 500 venues is intentionally set to be relatively narrow, covering high impact venues for service evaluation purposes; Places that are not within the GSR corpus should not be assumed to be of low quality.

### 5.3 Policy Suggestions

The cessation of CAS partitioning in March 2026 [3] poses immediate practical issues for institutions that have embedded partitioning metrics into formal evaluation rules. GSR is designed as a viable alternative: fully open source, capable of annual updates, covering the most frequently cited journals under CAS constraints, and providing a strict quantitative basis for partitioning. We suggest that institutions transition to FWCI normalization systems instead of seeking homogeneous alternatives to CAS - the core flaw of CAS and JIF based systems is the use of cross domain incomparable original citation counts [8,9]. We also suggest that institutional policies clearly distinguish the evaluation criteria between CS venues and medical/life science venues, and adopt a unified journal conference ranking for CS evaluation, rather than default to using a journal exclusive system that is structurally disadvantageous to the field.

## 6. Conclusion

This article proposes the General Science Ranking (GSR), an academic journal and conference partitioning system based on open data and domain normalization. GSR simultaneously addresses the three fundamental gaps in existing ranking tools: reliance on proprietary data, inability to include conference proceedings, and unadjusted citation metrics in the field of use. Based on approximately 1.87 million CS papers and 3.83 million medical papers from OpenAlex [6] and Semantic Scholar, GSR provided empirically driven and reproducible partitioning results for 500 CS sites and 500 medical journals. The ranking is consistent with the recognized external validity criteria (84% compliance rate between medical journals and JCR Q1), while correcting for known biases in the original influencing factors [8,9]. All code, configuration, and ranking results are provided for free in open-source form on https://github.com/gsci-press/General-Science-Ranking, supporting independent replication and institutional adoption.

**Acknowledgments**

We thank OpenAlex and Semantic Scholar teams for providing open data access.

**Data and code availability**

All analysis code, configuration files, and ranking results are open sourced and published under the MIT license on
https://github.com/gsci-press/General-Science-Ranking.
Data collection relies solely on publicly accessible API: OpenAlex (https://openalex.org) and Semantic Scholar (https://api.semanticscholar.org). No proprietary databases were used.